\documentclass[12pt]{iopart}
\usepackage{graphicx, bm, color} 
\usepackage{siunitx}
\usepackage{amssymb}
\usepackage[bookmarks=false]{hyperref}
\usepackage{cases}

\usepackage{cite}
\begin{document}

\title[Optimizing Ion Acceleration using PIC Simulations and Evolutionary Algorithms]{Optimizing Laser-Plasma Interactions for Ion Acceleration using Particle-in-Cell Simulations and Evolutionary Algorithms} 

\author{Joseph R. Smith$^1$, Chris Orban$^{1,2}$, John T. Morrison$^{2}$, Kevin M. George$^{2}$, Gregory K. Ngirmang$^{3,4}$, Enam A. Chowdhury$^{1,5,6}$,  W. Mel Roquemore$^4$ }

\address{ $^1$Department of Physics, The Ohio State University, Columbus, OH \\
$^2$Innovative Scientific Solutions, Inc., Dayton, OH \\
$^3$National Academies of Sciences, Engineering, and Medicine, Washington DC, USA\\
$^4$Air Force Research Laboratory, Dayton, OH \\
$^5$Intense Energy Solutions, Inc., Plain City, OH\\
$^6$Department of Materials Science and Engineering, The Ohio State University, Columbus, OH}

\ead{smith.10838@osu.edu}

\vspace{10pt}
\begin{indented}
\item[]
\end{indented}

\begin{abstract}

The development of ultra-intense laser-based sources of high energy ions is an important goal, with a variety of potential applications. One of the barriers to achieving this goal is the need to maximize the conversion efficiency from laser energy to ion energy. We apply a new approach to this problem, in which we use an evolutionary algorithm to optimize conversion efficiency by exploring variations of the target density profile with thousands of one-dimensional particle-in-cell (PIC) simulations. We then compare this ``optimal" target identified by the one-dimensional PIC simulations to more conventional choices, such as with an exponential scale length pre-plasma, with fully three-dimensional PIC simulations.  The optimal target outperforms the conventional targets in terms of maximum ion energy by 20\% and show a significant enhancement of conversion efficiency to high energy ions. This target geometry enhances laser coupling to the electrons, while still allowing the laser to strongly reflect from an effectively thin target. These results underscore the potential for this statistics-driven approach to guide research into optimizing laser-plasma simulations and experiments.

\end{abstract}

%
%
%
%
%

\section{Introduction}

Ultra-intense laser-based sources of energetic ions hold great potential to compactify and make more widely available the technology needed to accelerate ions to many MeV energies and higher~\cite{hatchett2000tnsa,snavely_2000_tnsa,scalingfuchs2006laser,higginson2018near}. Recently, up to 2~MeV proton acceleration was demonstrated with a kHz repetition rate laser system at the Air Force Research Lab (Morrison et al.~\cite{morrison_etal2018}). In a subsequent perspectives article ``Paving the way for a revolution in high repetition rate laser-driven ion acceleration," Palmer~\cite{Palmer2018} comments on Morrison et al.~and prior studies~\cite{McKenna_2002,Thoss_2003,Karsch_2003,Ter-Avetisyan_2006,Hou_2009,Hou_2011,Hah_2016,Dover_2017,Noaman-ul-Haq_2017}, arguing that experiments have now reached the point where these high-repetition-rate laser systems can be explored ``to provide compact accelerators for research and industry." With the rapid advancement of laser technology, the capability to accelerate significant numbers of many MeV ions from a compact source is becoming feasible for a variety of applications, including  proton imaging \cite{pRad}, hadron therapy for cancer treatment \cite{Bulanov2002,Tian_etal2018} and materials science.

An important next step in the translation from proof-of-concept experiments to these applications is to achieve more control over the properties of the laser-accelerated ion beam. Due to the complexity of ultra-intense laser interactions, rather than explore the large simulation parameter space  essentially by hand or some other means, instead we use an evolutionary algorithm with a series of thousands of one-dimensional (1D) particle-in-cell (PIC) simulations to optimize the laser plasma interaction.   The wider field of plasma physics is beginning to embrace statistical methods for various problems such as inertial confinement fusion \cite{gopalaswamy2019tripled,Rose2019,hatfield2019Gaussian,hsu2020NIFML}, magnetic fusion \cite{miner2001use,baltz2017achievement},  x-ray production~\cite{streeter2018temporalxray}, laser-wakefield acceleration \cite{he2015coherent,Lin:19}, and to optimize the laser focus for electron or ion acceleration experiments~\cite{Nayuki_tape_GA,Feister_2017}. To our knowledge, the present study is the first to directly optimize laser-based ion acceleration with such an approach.


Increasing the peak energy of the ions, while important, is only one of the properties of the ion beam that needs to be improved. In this paper we consider what can be done to increase the conversion efficiency between short-pulse laser energy to energetic ions ($\gtrsim 3$~MeV). 
In particular we explore ion acceleration using different target density profiles, while keeping the laser parameters fixed, in thousands of 1D PIC simulations as described in \sref{sec:1Dsetup}. Then, in \sref{sec:3Dsims}, we describe results from a few three-dimensional (3D) PIC simulations. These simulations allow us to more realistically examine whether the optimum target density profile found from 1D simulations in \sref{sec:1Dsetup} will indeed enhance ion acceleration compared to more conventional targets. Our results point to a novel type of target for enhancing ion acceleration, showing the potential of this method as discussed in \sref{sec:discussion}.

\section{1D PIC Optimization Driven by Evolutionary Algorithms}\label{sec:1Dsetup}
Evolutionary algorithms are a broad class of metaheuristics inspired by the biological theory of evolution\cite{holland1962outline,Goldberg:1989:GAS:534133,eiben2003introduction,coello2007evolutionary}. Within this class are ``genetic algorithms" and in this study we use an evolutionary algorithm called ``differential evolution"~\cite{Storn1997,bookprice2006differential}, which is specifically designed to deal with continuous variables. Evolutionary algorithms seek to optimize a `fitness function' (or `objective function') by testing many different candidate hypotheses creating a `population'  that reproduces and evolves over many generations~\cite{mitchell1997machine}.   For our work, the population is composed of many 1D PIC simulations and the `genome' represents the search space, where each `gene' is a parameter corresponding to one density bin throughout the depth of the ten-dimensional target density profile. Our specific implementation is discussed in detail in \ref{ap:DE}.

This approach allows us to explore a large parameter space in a highly parallelizable way, but there are several limitations that must be kept in mind. First, evolutionary algorithms are not guaranteed to find the global maximum and simulation choices further restrict the search space. Second, in order to quickly perform simulations we use 1D(3V) PIC simulations (one spatial dimension and three particle velocity dimensions) that are known to not be as realistic as 2D or 3D PIC simulations (e.g.~see \cite{Ngirmang_etal2016,stark2017effects} for differences between 2D and 3D simulations). Notably, 1D(3V) PIC simulations do not capture the focusing of the laser or the drop off of the electric field. To address this, we later present results from a 3D simulation that uses the optimal target from the 1D simulations. Despite these limitations, the prior success of evolutionary algorithms in related fields \cite{miner2001use,Nayuki_tape_GA,he2015coherent,Feister_2017,streeter2018temporalxray,Lin:19,Rose2019} demonstrates the power of this approach, and in this work, we find it to be advantageous for optimizing ion acceleration.


For this work,  we use a population size of 120 and let the algorithm evolve for 50 generations\footnote{The population size was chosen for computational and methodological considerations as discussed in \ref{ap:DE}, and the number of generations was based on convergence results (\sref{sec:1dparam}).}, resulting in a total of 6,000 simulations. Each simulation ran on a single core of a 2.4 GHz (Intel Xeon 6148) 20 core processor for approximately 30 minutes, resulting in a total execution time of about 2 days for all 50 generations. By starting with 1D simulations, we explore orders of magnitude more target configurations than would be possible with two-dimensions with the same spatial and temporal resolution.

\subsection{Simulation Parameters}\label{sec:1dparam} 
There are now more than one-hundred ultra-intense laser facilities in the world~\cite{icuil}. Rather than simulate some futuristic laser system, we chose to model a laser similar to the kHz repetition-rate laser described in Ref.~\cite{morrison_etal2018} with an 800~nm wavelength, $1.2\times 10^{19}$~W~cm$^{-2}$ peak intensity, and 42~fs full width at half maximum (FWHM) pulse duration. Electrons oscillating in these laser fields will experience significant relativistic effects and this intensity is sufficient to accelerate ions using the Target Normal Sheath Acceleration (TNSA) mechanism \cite{hatchett2000tnsa,snavely_2000_tnsa, Mora2003,Wagner_etal2006} and potentially other acceleration processes.

\begin{figure}
\centering
\includegraphics{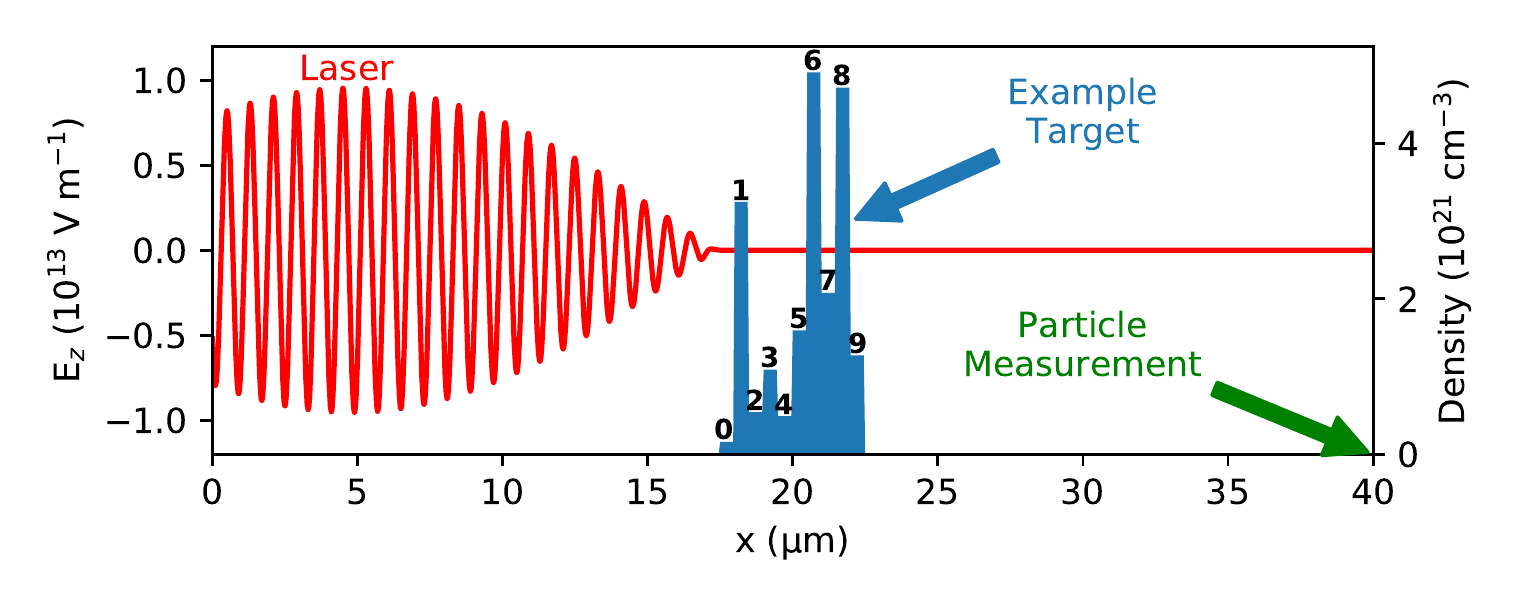}
\caption{Template for each of the 1D PIC simulations run by the evolutionary algorithm. The intense $1.2\times 10^{19}$~W~cm$^{-2}$ laser enters the simulation from the left side of the simulation box ($x$ = 0~\si{\um}) and interacts with a 5~\si{\um} thick ionized hydrogen target composed of ten 0.5~\si{\um} thick density bins chosen by the evolutionary algorithm. Ions and electrons are measured as they leave the right side of the simulation box ($x$ = 40~\si{\um}). The total energy of these ions is maximized with the evolutionary algorithm.     }  \label{fig:setup}
\end{figure}

\Fref{fig:setup} illustrates the blueprint of the 1D(3V) implicit PIC simulations run with the LSP PIC code~\cite{Welch_etal2004}.
The laser enters the  40~\si{\um} wide simulation box at $x=0$ and propagates towards a 5~\si{\um} thick target density profile that is generated by the evolutionary algorithm for each simulation. The target is composed of fully ionized hydrogen (protons and electrons) for simplicity and computational speed. As shown in \fref{fig:setup}, the density profile has ten independent 0.5~\si{\um} thick density bins. Although there is important work being done to create new kinds of targets for high-repetition-rate laser systems \cite{poole2014liquid,morrison_etal2018,koralek2018liquidTarget,george2019high}, we did not limit the search based on the current practicalities of what kinds of density profiles can be made in the lab\footnote{Various plasma shaping techniques will also facilitate new high-repetition-rate targets (e.g.~\cite{suntsov2009femtosecond,Lehmann2019,smith2019}).}. These bins are initialized randomly, by sampling from a uniform distribution, with a density up to $5\times 10^{21}$~cm$^{-3}$. For comparison, the classical critical density for laser propagation is $n_{\rm crit} = {4\pi^2\varepsilon_0 m_e c^2}/{\lambda^2e^2}$, or $1.7 \times 10^{21}$ cm$^{-3}$ for an 800~nm laser. For high intensity lasers, relativistic effects increase this density to $\gamma n_{\rm crit}$ where $\gamma$ is the Lorentz factor for the electrons, but for the intensity we consider here the relativistic critical density is still generally below $5 \times 10^{21}$~cm$^{-3}$.

To optimize the conversion efficiency from laser energy to ion energy, the fitness function was the total energy of ions that leave the right edge of the simulation boundary (\fref{fig:setup}). This choice ignores the backwards (i.e. leftward) going ions. Also, due to the finite simulation time, ions with less than $\sim 3$~MeV may not reach the right edge of the simulation box. As the algorithm evolves the targets, we do not limit the maximum density to $5 \times 10^{21}$~cm$^{-3}$ and instead allow the densities to grow above this value with no upper bound. If the evolutionary algorithm selects a negative density value for one of the density bins, that density is set to zero. While it is generally advisable to allow the initial random parameter selection to span the entire search space, we found that this skewed the initial population to significantly overdense targets that were not as conducive to ion acceleration.

The simulations have a spatial resolution of 10~nm ($\lambda/80$) with 64 particles per cell for each species.  The macroparticles are initialized with thermal temperatures of 1~eV and the simulation time is 1,000~fs with a time step of 0.05~fs. The spatial scale does not resolve the Debye length for all possible target configurations, however the implicit field solver and energy conserving algorithms of LSP limit artificial grid heating. Collisions are allowed in the code, although turning off collisions does not make a significant difference when tested with the optimal conditions in 2D simulations. 

\subsection{1D Results}
\begin{figure}
\centering
\includegraphics{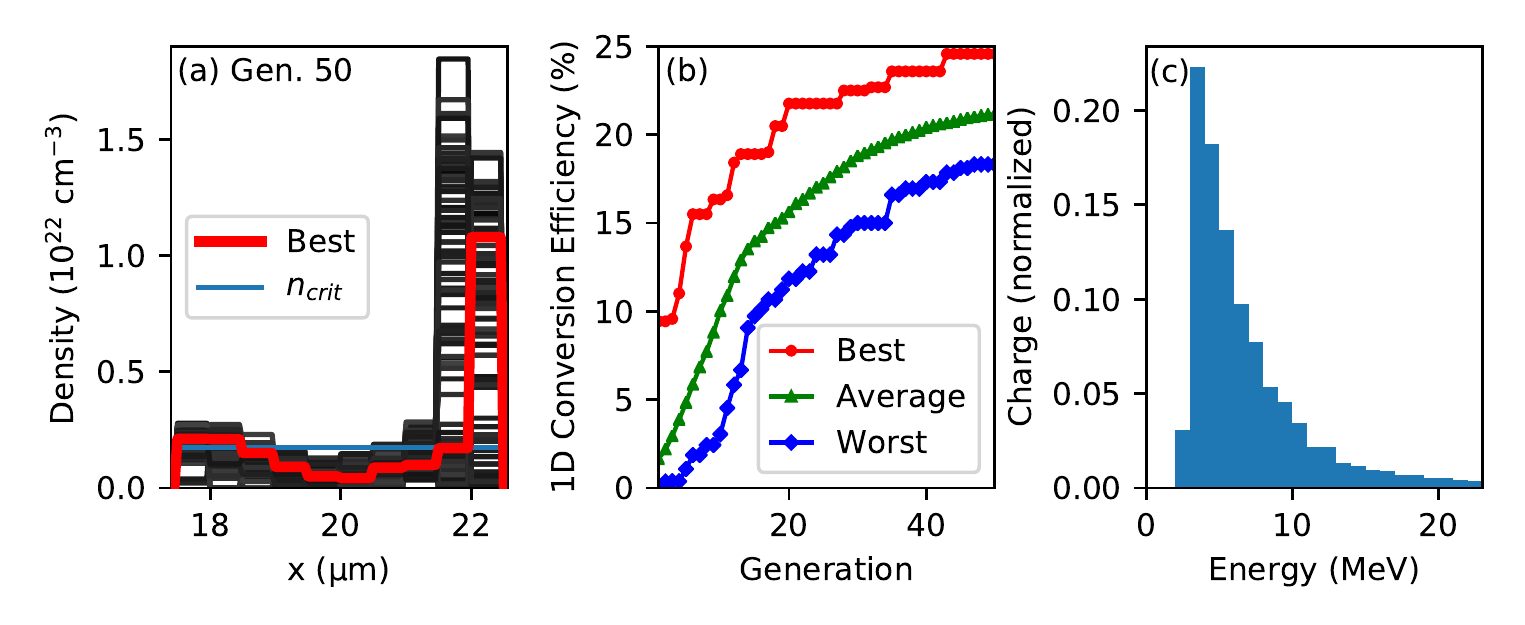}
\caption{1D PIC simulation optimization results. In (a), all members of the population are plotted, where darker shades represent higher fitness. The best performing density profile is drawn in red. After 50 generations, most members of the population have a similar pattern with a roughly critical density foot at the front of the target (the first two density bins that exceed $n_{\rm crit}$), underdense center, and overdense density spike in one of the last two density bins as shown.  In (b), the conversion efficiency of measured ions initially increases quickly and then begins to level off with later generations. In (c), the distribution of measured forward going ions for the best performing profile is plotted. This `optimal' density profile is tested with 3D simulations in \sref{sec:3Dsims}.       }  \label{fig:Conv}
\end{figure}

 \Fref{fig:Conv}(a) shows the population after 50 generations, which converges to a general density profile.  Density profiles from earlier generations are presented in \ref{ap:1D}.
For the optimal density profile shown in \fref{fig:Conv}(a) (drawn in red), there is a classically overdense foot at the front of the target for the first two density bins. This is followed by classically underdense bins in the center of the target and an overdense spike in the last density bin. The dark lines in \fref{fig:Conv} show that all of the other members of this final generation follow a similar trend with reduced density in the center of the target and an overdense spike one of the last two density bins.

\Fref{fig:Conv}(b) shows how the conversion efficiencies of the simulations improve with each generation. In the initial generation, the best performing target has close to 10\% conversion efficiency. The following generations improve upon this result, eventually reaching nearly 25\% in the 50th generation, with most of this improvement coming from the first ten or so generations. The last ten generations only improve the conversion efficiency by a small amount, which is part of our rationale for ending the generations at 50. \Fref{fig:Conv}(c) shows for the optimal target (generation 50) the distribution of energies for ions ejected from the target. The highest energy ions exceed 20~MeV, which is quite high for these laser parameters, but the approximations made by 1D simulations typically overpredict the maximum energy so we will hold our comparison to typical targets until the higher dimensional simulation results are presented.

\section{Three-Dimensional Simulations}\label{sec:3Dsims}
\begin{figure}
\centering
\includegraphics{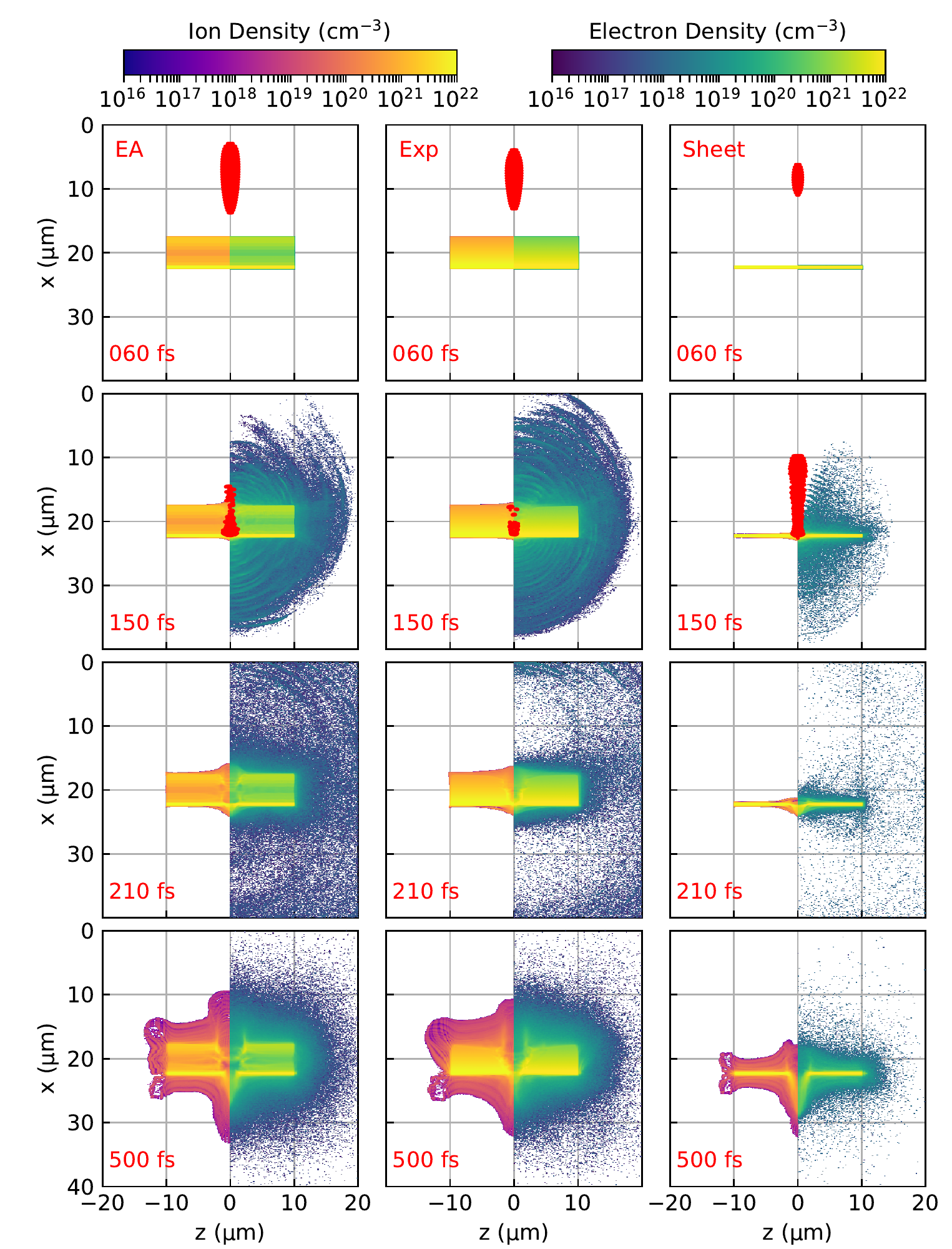}
\vspace{-1em}
\caption{Snapshots of the ion ($z < 0)$ and electron ($z>0$) densities for the three 3D simulations ($xz$ plane). The optimal `evolutionary algorithm (EA)' target from the 1D simulations is on the left, the exponential target is in the center, and the thin sheet is on the right. Ions travel farthest for the new EA target, which shows enhanced coupling between the laser and electrons like the exponential target. A contour is drawn at an intensity of $I_{max}/e^2$; variations come from differences in the focal spot location.  See the supplementary material for an accompanying video.  }  \label{fig:3Devolution}
\end{figure}

 To better understand this new type of target identified with the 1D PIC simulations, we performed a series of 2D(3V) and 3D PIC simulations. For brevity, we focus on the results of the 3D simulations for a 4 $\times$ 10$^{19}$ W cm$^{-2}$ laser with the longitudinal density profile of the best target from the evolutionary algorithm, represented with a 20~\si{\um} wide target. We compared these simulation results to the more conventional targets of a thin 0.5~\si{\um} sheet (with the same density as the last density bin of the evolutionary algorithm target) and a target with a 1.5~\si{\um} exponential scale length in front of the sheet. A 2D slice of all three targets is shown in \fref{fig:3Devolution}.

\subsection{Simulation Parameters}
The simulation parameters for the 3D simulations closely match those of the 1D simulation in \sref{sec:1dparam} to test whether the same behavior persists in higher dimensional simulations.  The 3D simulations were conducted with $4\times 10^{19}$~W~cm$^{-2}$ lasers so more interesting ion energies could be explored, while staying on the frontier of capabilities of current kHz laser systems.

Earlier in \sref{sec:1dparam} we did not specify a spot size or the position of peak focus for the laser pulse because focusing is not accounted for in 1D PIC simulations. For the 3D simulations, we assume a Gaussian spot size of 1.5~\si{\um} (FWHM) and we set the peak focus at the front of the target ($x = 17.5$~\si{\um}) which allows most of the laser pulse to propagate through the classically overdense section of the target there via relativistic transparency. For the exponential scale-length target, the focus was set near the critical density at $x = 19.3$~\si{\um} and for the sheet it was set at $x = 22.5$~\si{\um}. The spatial resolution of these simulations is 50~nm ($\lambda/16$) in the laser propagation ($x$) direction and 100~nm ($\lambda/8$) in the transverse directions and 125 particles per cell were used for each species. This is a much lower resolution than the 1D simulations, although 2D tests indicated that these conditions are sufficient to model the process. Despite this lower resolution, one 3D simulation required over 30 times more computational resources than all 6,000 1D simulations. 

\subsection{Results}
\Fref{fig:3Devolution} shows snapshots of the ion and electron densities at several points in time throughout the three simulations. We observe that the ions are able to travel noticeably farther away from the back of the target in the simulation with the evolutionary algorithm target, suggesting higher maximum energies than the conventional targets, as explored shortly. For the evolutionary algorithm and sheet target, the laser does not reach the critical density until it reflects at the last density bin near the back of the target allowing the laser to interact with an effectively thinner target. The sheet target has many fewer electrons expanding from the target than the other two simulations as shown \Fref{fig:3Devolution}.  The total electron energy during the simulation rises to a maximum of 44\% of the total incident laser energy for the evolutionary algorithm target, 56\% for the exponential target and only 12\% for the sheet target.

\begin{figure}
\centering
\includegraphics{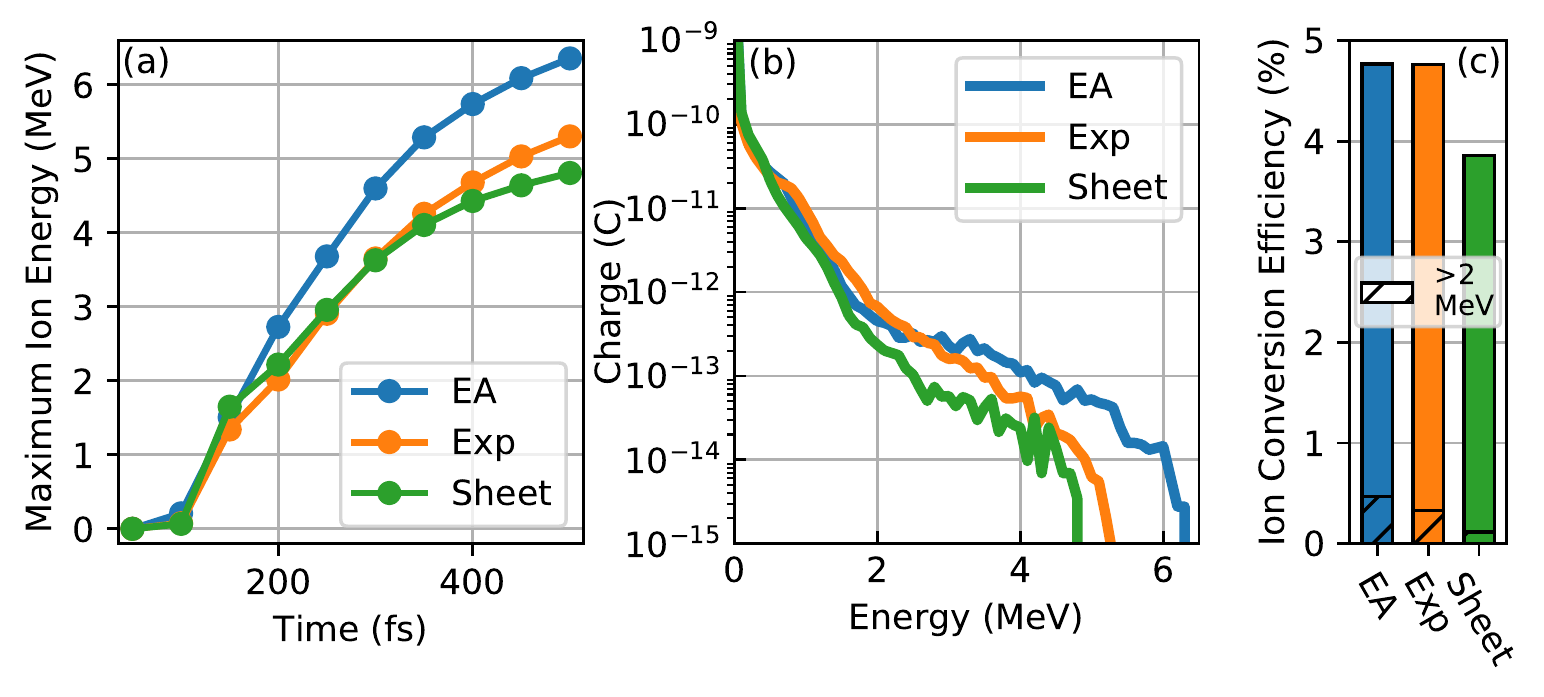}
\caption{ The maximum ion energy versus time for all three simulations (a), where the EA target reached the highest energy followed by the Exp and Sheet targets.  Spectra of ions with forward going momenta in a 20$^\circ$ half angle cone sketched in \fref{fig:Polar}, for the three different 3D simulations at 500~fs (b), where the charge represents the total charge of electrons in a 0.1 MeV energy bin. The total ion conversion efficiency in this cone is included in (c) with the hatched bars showing the difference between the targets if only ions with energies greater than 2 MeV are considered. The overall conversion efficiency is similar, but there is a noticeable enhancement to the population above $\sim$2~MeV for the EA target compared to the other two targets.  }  \label{fig:spec3D}
\end{figure}

\begin{figure}
\centering
\includegraphics[width=\linewidth]{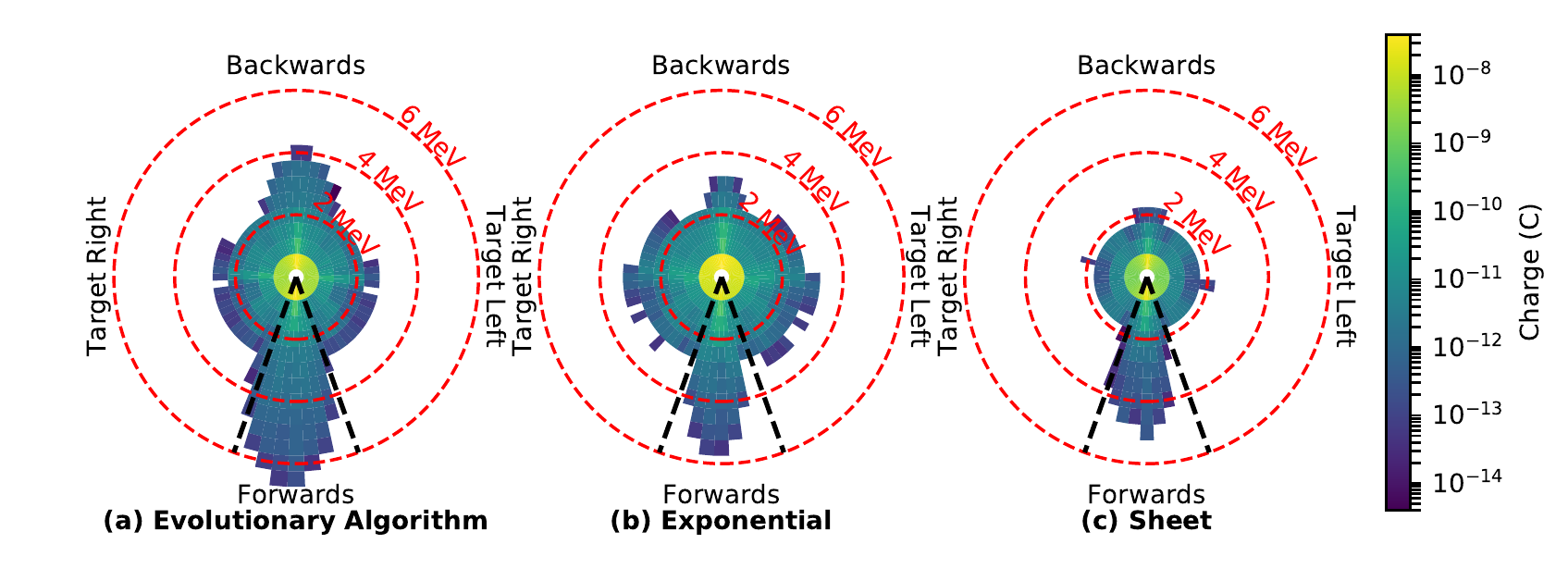}
\caption{Polar histograms showing the distribution of ion energies for the three 3D simulations. Energy bins have a radial size 0.5~MeV and angular size of 5$^\circ$ taken in the $xz$ plane and a $20^\circ$ half-angle cone is sketched for reference. For the evolutionary algorithm target (a), we see a strong forward (laser propagation direction) going component of the ion distribution and enhanced conversion to $\gtrsim 2$~MeV ions compared to the other two targets.  }  \label{fig:Polar}
\end{figure}

\Fref{fig:spec3D}(a) shows the maximum ion energy versus time for each simulation.
Around 150~fs, the evolutionary algorithm target begins to outperform the other targets in terms of maximum ion energy and for later times exhibits sustained growth similar to the exponential target for the rest of the simulation. The targets have maximum ion energies of about 4.8 MeV for the sheet, 5.3 MeV for the exponential target, and 6.4 MeV for the evolutionary algorithm target.  We compare the spectra of forward going ions at the end of each simulation in \fref{fig:spec3D} for the $20^\circ$ half-angle cone sketched in \fref{fig:Polar}, which shows that higher ion energies are obtained with the new target geometry. 
From \fref{fig:spec3D}(b), we see that the exponential target has a slightly larger population of ions at lower energies, but drops off more quickly for higher ion energies. The higher energy ions are of interest for many applications~\cite{Fritzler_2003,Ledingham_2004,ene2009applications,jain2011ion}. As illustrated in \fref{fig:spec3D}(c), the exponential and evolutionary algorithm targets had similar overall conversion efficiency in this cone of about 4.8\%, but there were significant differences when considering higher energy ions. For example, conversion efficiency to $>$ 2 MeV ions is 0.47\% for the evolutionary algorithm target, 0.33\% for the exponential target and 0.12\% for the sheet target.

As shown in \fref{fig:Polar}, the highest energy ions are traveling in the forward (laser propagation) direction and are primarily contained within a $20^\circ$ half-angle cone. While not the focus of this work, \fref{fig:Polar} also shows more significant back directed ions from the evolutionary algorithm target than the other cases. One can also see that the exponential target has more significant semi-isotropic ion acceleration (i.e.~at large angles) than the other targets, which is not surprising due to the higher laser-electron coupling.


\section{Discussion}\label{sec:discussion}
\subsection{The Optimal Target}
In this subsection we comment on the distinct features of the optimal 1D target (\fref{fig:Conv}(a) shown in red) that make it an interesting new candidate for ion acceleration. There are three basic elements of the optimal target: classically overdense foot, near-critical density cavity, and an overdense spike. The classically overdense foot at the start of the target becomes relativistically transparent, allowing a majority of the laser pulse to pass. This first phase of the interaction is reminiscent of studies where ion acceleration is enhanced by relativistic transparency (e.g.~\cite{Wagner_etal2006,higginson2018near,Poole_2018}). Next, the laser propagates through the near-critical density cavity, transferring significant energy to electrons. However, unlike the targets in the aforementioned studies, the laser reaches an overdense spike where it makes a strong reflection because the density of the spike significantly exceeds the relativistic critical density. The outgoing pulse continues to transfer significant energy to electrons as it passes through the near-critical density cavity a second time. Then the pulse reaches the foot of the target and escapes. In the 1D simulations, some of the laser pulse appears to become trapped in the cavity, although this effect is not significant in the 3D simulations, which is likely due to 3D considerations such as the focusing of the laser pulse.

Through the use of thousands of 1D simulations and an optimization method, we identified a new type of target, not yet explored with experiments, that employs commonly studied laser plasma effects. Because of the strong reflection, the ``optimal" target is comparable to efforts that use reflection to better confine the laser energy and enhance coupling. So-called plasma half cavity targets use a hemispheric reflecting surface to direct laser light back to the interaction region \cite{Scott_2012}. So-called ``escargot" targets use reflections to direct the laser light into a kind of spiral \cite{Korneev_sn_PhysRevE}. Both of these approaches involve much larger scales than the few-micron thick targets we consider here.   
On smaller scales, two studies that consider electron heating in high-reflectivity laser interactions are \cite{Kemp_etal2004} and \cite{Orban_etal2015}. Although there is both constructive and destructive interference where the laser pulse overlaps, the constructive interference can enhance the population of hot electrons, which is well known to play an important role in TNSA. Recent simulation work with nanostructured double-layer targets, such as a random forest of nanowires on a thin target, shows enhanced ion acceleration due to increased laser-to electron coupling like our work~\cite{fedeli_2020_DLT}.

\subsection{Optimization of Laser Plasma Interactions}
We searched a 10-dimensional parameter space, considering 5~\si{\um} thick targets with rather course 0.5~\si{\um} thick density bins and using a single laser intensity and pulse duration. Even with this limited search space, we identified a new type of target that seems to match or outperform conventional targets in terms of maximum ion energy and conversion efficiency to higher energy ions. 
There is still a vast parameter space of laser plasma interactions to be explored with this technique and others. We have also generated a data set of 6,000 simulations that is being examined to find additional trends\footnote{The parameter search is inherently biased by the evolutionary algorithm, but still reveals some interesting features.}. This proof of concept illustrates the potential benefit of using many 1D simulations to discover new target geometries for laser plasma interactions. This method is not limited to ion acceleration may be useful for tasks such as optimizing pulse shapes for inertial confinement fusion~\cite{hurricane2014fuel}.

\section{Conclusions}
\label{sec:conclusions}

With the small computational cost of one-dimensional simulations and an optimization routine utilizing evolutionary algorithms to run thousands of 1D simulations, we identified a new type of target for enhancing ion acceleration. This new target was then examined with 3D simulations and showed enhancement compared to conventional targets.

One limitation of this approach is that 1D PIC simulations are much less realistic than 2D or 3D simulations. Future efforts using large numbers of 1D PIC simulations to guide efforts to optimize laser plasma interactions may not be as generally successful as was demonstrated in our study. We noticed, for example, that 1D simulations saw the ``trapping" of the laser pulse due to a second reflection inside the cavity inside the target whereas this phenomenon was not seen in 3D simulations. There are also targets with complicated geometries that are not amenable to running 1D simulations (e.g.~\cite{ji2016tube}).

This work highlights the potential for using evolutionary algorithms and other statistical methods to study laser-plasma interactions in both experiment and simulation. There are many outstanding challenges in this field that may benefit from this approach such as increasing the maximum ion energy with current laser systems and efforts to produce monoenergetic ion beams.

\ack
This research is supported by the Air Force Office of Scientific Research under LRIR Project 17RQCOR504 under the management of Dr.~Andrew Stickrath and Dr.~Riq Parra. This project also benefited from a grant of time at the Onyx supercomputer (ERDC) and the Ohio Supercomputer Center~\cite{OhioSupercomputerCenter1987}. Support was also provided by the DOD HPCMP Internship Program. Some figures in this paper were generated using matplotlib~\cite{Hunter:2007}. J.~S. would like to thank Douglass Schumacher for insightful conversations about the work.

\appendix
\section{Differential Evolution}\label{ap:DE}
\begin{figure}
\centering
\includegraphics{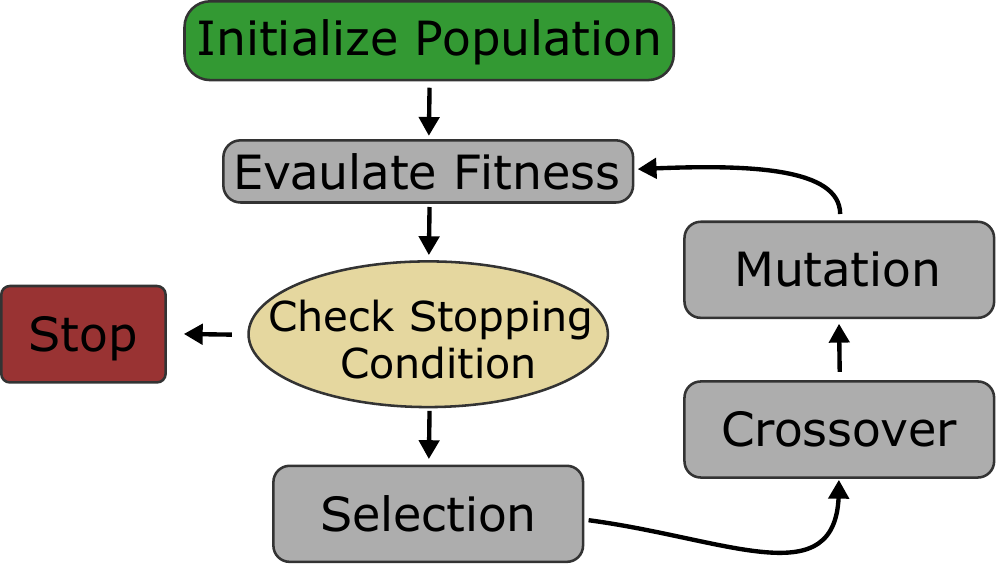}
\caption{The general procedure for an evolutionary algorithm. A population is created and evolved with crossover and mutation, based on a fitness function until a stopping condition is met. For our work, the `Evaluate Fitness' step depends on the output of 1D PIC simulations.    }  \label{fig:EA_Daig}
\end{figure}

The general procedure for an evolutionary algorithm is sketched in \fref{fig:EA_Daig} and is explained in depth in Refs.~\cite{Goldberg:1989:GAS:534133,mitchell1997machine}. We begin by initializing the population, typically a random sampling of the search space. Then the fitness of each member of the population is evaluated. If the maximum fitness of the population is within some threshold, or if it has reached the maximum number of iterations, the algorithm is complete. Otherwise, we proceed to selection, where the `parents' of the new generation are selected based on their fitness (the initial population may be used in whole as the first parents). Next `crossover' occurs, where two or more parents are mated to form a `child'. Then mutation occurs, where the genes of some children are modified. Finally, we evaluate the fitness of the children and the process repeats until the stopping condition has been satisfied. The stopping condition is triggered if the fitness reaches some predetermined value. In practice, if there is no stopping condition selected, the user may manually stop the evolution based on performance.  

For a full description of the differential evolution algorithm, we refer the reader to Refs.~\cite{Storn1997,bookprice2006differential}; but we provide a summary of the process here.  In differential evolution, four parents are used in the crossover/mutation steps, where to create the mutation vector, the parameter from one of the parents is perturbed based on the difference between the value of two other parents.  The algorithm begins by initializing $NP$ (population size) $D$-dimensional vectors randomly sampling the parameter space, which we will call  $\bi{x_i}$ where $i = 1,2,\dots NP$. In our case, we use 10-dimensional vectors, where each element corresponds to part of the density profile of a target (see \fref{fig:setup}). Then to find the test vectors for the next generation, we loop through all members of the population. For each $i$, we generate a mutant vector  
\begin{equation}
    \bi{m_i} = \bi{x_{n_1}} + F\cdot(\bi{x_{n_2}} - \bi{x_{n_3}}),
\end{equation}
where $\bi{x_{n_1},x_{n_2},x_{n_3}}$ are mutually distinct members of the population (also distinct from $\bi x_i$ which is used in \ref{eq:testVector}), and $F\in [0,2]$ is a factor that controls the weighting of the differential evolution~\cite{Storn1997}. Then to form the test vector $\bi{t_i}$ for the next generation, we select a crossover rate $CR \in [0,1]$. Next for each gene the crossover rate represents the chance that a gene is selected from $\bi{t_i}$. To do this, we generate loop over the genes $j = 1,2,\dots D$ and generate a random number between 0 and 1 ($\text{Rand}[j]$) to see determine crossover occurs, or


\begin{numcases}{t_i[j]=}
   m_i[j], & if  Rand[j] $\le$  CR\nonumber\\
   x_i[j], & otherwise. \label{eq:testVector}
\end{numcases}

If no genes have been selected from $\bi{m_i}$, one is automatically chosen to prevent testing of the same point twice (some implementations of the algorithm automatically switch one gene). The fitness is then calculated (a 1D PIC simulation is run in our case) and if the fitness of $\bi{t_i}$ is better than ${\bf{x_i}}$, it becomes a member of the next generation. We use $F =0.5$, and $CR=0.9$, which are initial parameter choices recommended by \cite{Storn1997}. Often a population size of ten times the dimension size is used \cite{Storn1997}, we slightly exceeded this with $NP=120$, which was a convenient choice as the computer system used had 40 cores per node. While typical values proved to have good performance for our problem, they can require significant tuning in practice, which would be an important consideration for problems with higher computational (or experimental) costs.  For our work, by far the largest computational expense in this process is the 1D PIC simulations (which determines the fitness of each member of the population). The mutation of the population and selecting from these mutations to create a new population requires negligible computational time by comparison.

We implemented the evolutionary algorithm in Python.  It selects the density profiles and then creates data files that are read by the PIC simulation code LSP. The simulation runs are initiated directly from the Python code with a system call to run an compiled LSP executable.\footnote{For increased computational demands, or different allocation structures, it may be beneficial to submit the PIC simulations to run as a separate job files. } Following the completion of all PIC simulations, the Python script reads the output files from LSP to calculate the fitness.    

\section{1D Evolution}
\label{ap:1D}

\begin{figure}
\centering
\includegraphics{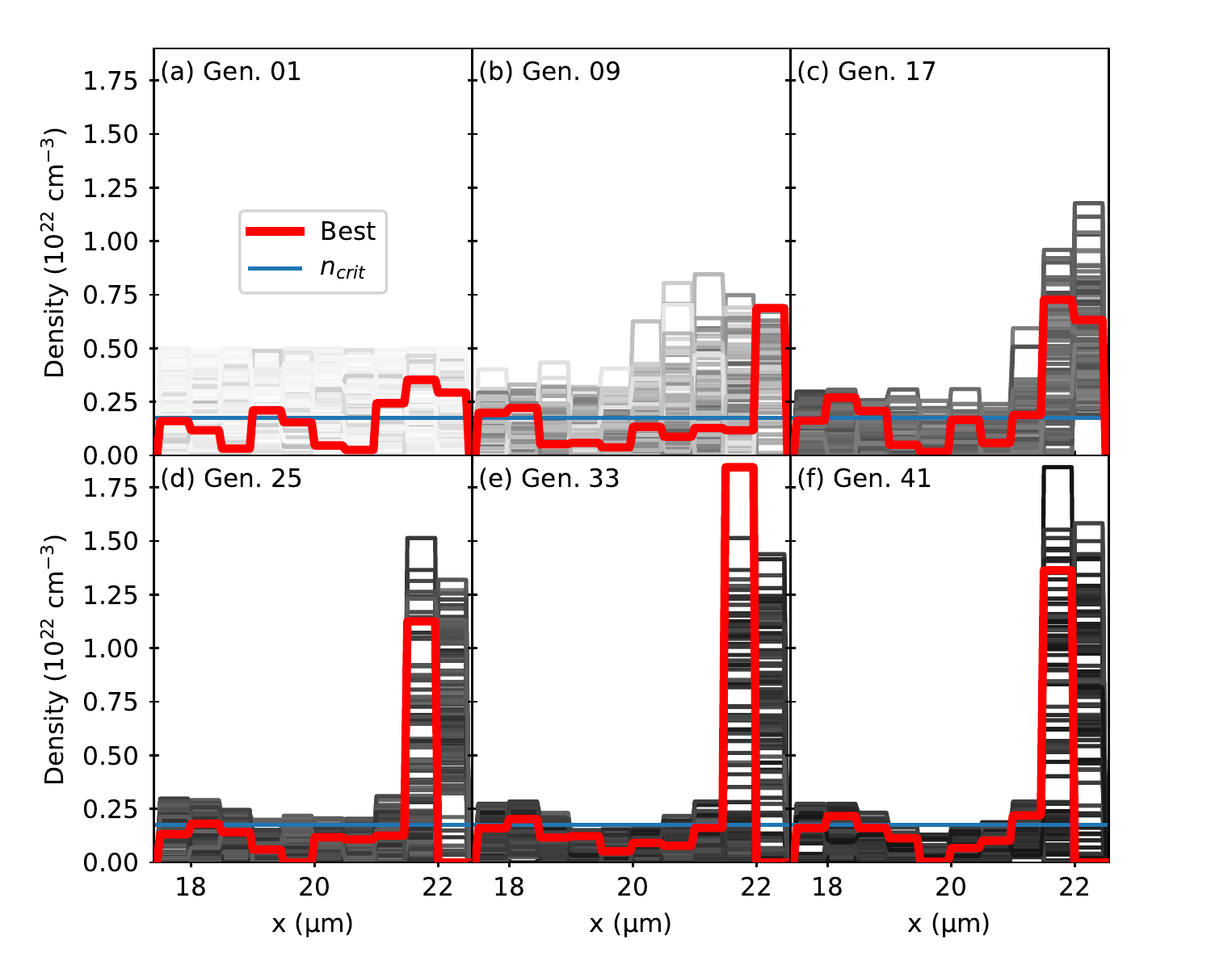}
\caption{Evolution of the population (a-f). The entire population is plotted, with darker lines indicating higher conversion efficiencies. The highest performing member of each generation is plotted in red and the classical critical density is plotted for reference.   }  \label{fig:1DEv}
\end{figure}

\Fref{fig:1DEv} includes additional snapshots of the population throughout the evolutionary algorithm. For the initial generation, we see a relatively uniform sampling of the parameter space, but with further generations patterns begin to emerge and the population becomes more homogeneous. After twenty or so generations, the front of the target is typically classically overdense, then the center of the targets in the population are primarily underdense and there is typically an overdense spike in one of the last two density bins.  

\bibliography{main}
\bibliographystyle{unsrt}

\end{document}